\documentclass[a4paper,11pt]{article}
\usepackage{pos}

\title{Theory Closing Talk}

\author*[a]{Gian F. Giudice}

\affiliation[a]{CERN, Theoretical Physics Department\\
  Geneva, Switzerland}

\emailAdd{gian.giudice@cern.ch}

\abstract{In this talk I make some remarks about the search for the origin of the electroweak scale.}

\FullConference{%
 
 The Ninth Annual Conference on Large Hadron Collider Physics - LHCP2021\\
 7-12 June 2021\\
 Online
}

\begin{document}
\maketitle

A crucial lesson learned from the first 190 fb$^{-1}$ of LHC data is that hadron colliders are not only discovery machines but also excellent precision machines. This result could not have been anticipated at the time the LHC started and was possible only because of the successful interplay between different elements: unprecedented technological advancements, exceptional accelerator performances, excellent detector resolutions, high-performance computing and data handling, higher-order theoretical calculations of background processes with accuracies unthinkable only a few years ago. The confluence of different expertise from different scientific communities was the secret behind the success of the LHC precision programme, which brought new knowledge and opened new prospects in research beyond traditional frontiers. Precision has become key for present and future exploration in high-energy physics.

Indeed today it is from the precision programme that we have received the most enticing news from the LHC: the $B$ anomalies from LHCb. If true, this result would be the most revolutionary surprise in fundamental physics since the discovery that the universe is expanding at an accelerating rate. The $B$ anomalies cannot be explained by a small SM deformation, as they really shatter the SM flavour structure and the symmetries on which it rests. If true, this result would radically change our perspective for future experimental activities, boosting the motivations for the flavour programme, the precision programme and the high-energy programme as well, because it would give us an indication for the mass scale of new physics. 

The information from the $B$ anomalies goes straight to the heart of one of the most puzzling open questions in particle physics: understanding the pattern of quark and lepton masses. Often particle physicists lament that they don't understand 95\% of the energy budget of the universe, which is in the form of dark matter and dark energy. The truth is that we don't even understand the other 5\% because we cannot make sense of the observed pattern of masses of the particles that form ordinary matter. We have made elaborate attempts to compute the Higgs mass, but we still have no clue on how to compute the mass of the electron, a particle which was discovered more than  100 years before the Higgs. If true, the LHCb data could give a clue as to how to attack the problem. 

Fermilab has added gasoline to the fire of the LHCb anomalies by confirming a discrepancy in the $g$--2 of the muon. Although this process doesn't involve flavour transitions, it is not impossible that the various anomalies, which share muons as common elements, may have a related origin. Luckily we will soon have a lot more information from experimental data (LHC, Muon $g$--2 @ Fermilab, and Belle II) and theoretical calculations (especially lattice results on $g$--2). It is too early to tell how the story will evolve, but certainly these are interesting times for particle physics.

Leaving aside these intriguing hints, the question that crosses every particle physicist's mind is why the LHC hasn't detected any direct evidence for new physics yet. Addressing this question puts us in front of a conceptual crossroads. One direction follows the traditional approach, which I will call the {\it symmetry paradigm}. 

The Standard Model and General Relativity combined together are probably one of the greatest scientific achievements of humanity. They are based on the symmetry paradigm, in the sense that these theories are fully determined by their gauge symmetries, once the matter content is specified. Following the road of the symmetry paradigm to explain the origin of the EW scale has led us to breathtaking new concepts that imply new particles, new types of forces, new symmetry structures, a new vision of spacetime. But we haven't seen any of the new phenomena associated with these new concepts. This doesn't conclusively exclude the symmetry road for the origin of the EW scale because there could be a slight scale separation between new and SM phenomena. Indeed, this happens in several known theoretical models, which are logically consistent and have solid scientific justifications. Now the question is purely experimental. We have the tools to pursue the symmetry direction and we should pursue it. This is a well-motivated research direction which will provide us with new fundamental knowledge, no matter what the data will show. However, in spite of the extraordinary successes of the symmetry paradigm in the past, there are hints that the concept of symmetry may be running out of gas. 

Experimental indications come from the absence of new phenomena at the LHC and from the evidence for a cosmological constant. Indeed, from an EFT point of view, the naturalness problems with the Higgs mass and the cosmological constant look fundamentally identical. A difference is that the cosmological constant becomes an observable only after you turn on (classical) gravity, but this is unimportant for our considerations.  

There are also theoretical indications for questioning the concept of symmetry. It is now believed (and to a certain extent proven) that any global symmetry is violated at the level of quantum gravity. This means that any global symmetry that we observe in nature is only an accidental effect of looking at a system without sufficient short-distance resolution. The case of gauge symmetries is more subtle. Gauge symmetries are not real physical symmetries, in the sense that they don't correspond to an invariance under a physical transformation, but only to a redundancy of the coordinate parametrisation. We often confuse our students on this point by showing them the Mexican-hat potential and leading them to believe that there is a degeneracy of vacua, when in reality there is only one single vacuum state that breaks EW symmetry, as it is clear from the fact that the physical spectrum doesn't contain any Goldstone boson corresponding to zero-energy excitations.  Gauge symmetries may not be as fundamental as we thought, but only an emergent phenomenon. They could be a mirage of a different reality that takes place at a more fundamental level. Maybe the LHC is telling us that it is not just a matter of adapting our models or adjusting some parameters, but it is really time to look for radically different paradigms. 

When questioning a fundamental physical concept (such as naturalness), a good practice is to rely on the scientific method. In the scientific method, one starts from some hypotheses, makes a thesis, and then finds out if experiments confirm the thesis. If not, one starts questioning the hypotheses. So let's see what are the hypotheses on which naturalness rests.

The framing of the naturalness problem is based on the existence of multiple well-separated energy scales. If you take the SM in isolation as a renormalisable theory, you can't even formulate the problem because there is no hierarchy of scales to start with, at least at the classical level. When quantum effects are included, new mass scales seem unavoidable even in the SM alone, as signalled by the presence of Landau poles. Since scale separation is the first essential hypothesis for naturalness, the question is: does nature contain other energy scales above the weak scale? We have good reasons to believe that the most plausible answer is yes. Quantum gravity has a scale associated with Newton's constant. It is not a mass scale, but it is very plausible that the Planck mass corresponds to some new heavy modes. Other indications come from neutrino masses, the strong CP problem, inflation, gauge coupling unification, Landau poles, and the lack of a SM explanation for the pattern of quark and lepton masses. None of them can be taken as indisputable evidence, but their combination gives a strong indication for the existence of new heavy modes in nature. Nonetheless, the validity of this hypothesis is not guaranteed and it was challenged, for instance, by large extra dimensions, which tried to bypass naturalness by postulating that there is no scale separation in the real world, or by approaches inspired by asymptotic safety in quantum gravity. This is a very ambitious route because it assumes that all problems in the SM can be solved simultaneously by an ultimate theory defined at the weak scale but whose validity extends to all energy scales. In practice it is 
hard to reconcile the idea of no scale separation with the real world.

The second hypothesis on which naturalness rests is EFT validity or, in other words, that physical phenomena can be consistently described by an effective theory valid in a limited energy range. All information about short distances can be buried into some free parameters of the effective theory and the description of a low-energy phenomenon doesn't require knowledge of the final theory of nature, valid at all energy scales. This hypothesis is an essential tool for the physicist's job and has always been taken for granted. For centuries, physics has made progress exactly because natural phenomena can be described layer by layer. Newton didn't need to know about the atomic structure of matter to compute the orbit of the Moon around the Earth. Bohr could compute the atomic levels without knowing about the existence of quarks or gluons. We have become accustomed to the fact that we can give a consistent physical description within a certain energy range and physics can make progress proceeding step by step, without full knowledge of physics at arbitrary short distances. However, it is not guaranteed that nature satisfies this hypothesis.

The violation of the EFT Wilsonian intuition would be a complete revolution 
of our approach to particle physics. It is so revolutionary that we don't even know how to formulate a concrete example of how this could happen. This would require some magic interplay between IR and UV phenomena, introducing special correlations such that the total quantum correction to an IR parameter is much smaller than the single contribution coming from the UV, completely defying EFT logic and possibly requiring the breakdown of the concept of locality in QFT. This may sound totally crazy, but there are indications that an EFT may not catch the full physics story, even in the low-energy domain. The various swampland conjectures state that not all theories that are allowed by the usual rules of EFT, symmetry and selection rules, necessarily have a consistent UV completion in the context of quantum gravity. There could be more in a low-energy theory than simply the rules of symmetry and
there could be restrictions to the IR theory that are completely inexplicable from a low-energy point of view. Of course, this doesn't necessarily mean that IR/UV connections are possible, but it exposes the limitations of thinking only in terms of low-energy effective theories. 

Another intriguing consideration is that a QFT supplemented by Planck-suppressed higher-dimensional operators cannot consistently describe the low-energy behaviour of quantum gravity. The reason is that low-energy particle states can  collapse into black holes which, by Hawking radiation, eventually lead to regions where the curvature is Planckian and quantum gravity dominates the dynamics. Effectively, particle states evolve out of the EFT Hilbert space and this apparent violation of unitarity is a sign of an inner inconsistency of the EFT in presence of quantum gravity. It has been conjectured~\cite{Cohen:1998zx} that these considerations limit the EFT validity to a finite energy range in which the IR and UV cutoffs are related. This could be interpreted as another clue for a hidden IR/UV connection.

Dropping the EFT hypothesis could be an interesting adventure in the exploration of naturalness. However, it is a very difficult road to pursue because it necessitates blowing apart some of the founding principles of QFT. It is an interesting direction for bold mavericks with great ambition. 

The third hypothesis on which naturalness rests is that IR free parameters become calculable quantities in the UV completion. Whenever we formulate the question about Higgs naturalness, we tacitly assume that the Higgs mass is calculable in the fundamental theory. For instance, in supersymmetry the Higgs mass is given by a combination of the mass parameters that break supersymmetry and those that break a chiral symmetry. This is not surprising since the Higgs mass is protected by these two symmetries combined together. Similarly, the mass of the composite Higgs is given by a calculable expression proportional to the parameters that break the shift symmetry that would make the Higgs a Goldstone boson.  Naturalness implies that the mass parameters that control the relevant symmetries cannot be much larger than the Higgs mass itself. 

One trivial way of giving up this hypothesis would be to assume that the Higgs mass is not a calculable quantity, but a God-given number whose origin lies beyond human comprehension. Although this is a logical possibility, it is not a very scientific one. The history of science has taught us that there is a pattern in nature, that ``nature is comprehensible'' (to paraphrase Einstein) and that there are fundamental laws of nature for us to discover.

A more interesting way of conceiving that IR parameters cannot be simply calculated from the UV completion is to speculate that they are functions of some fields whose values vary during the cosmological history or throughout a complex vacuum structure realised in the universe. This point of view is a profound paradigm change with respect to the usual approach based on symmetries. 
 What we call SM parameters may actually be dynamical variables that take on different values in different parts of the universe, even well beyond our event horizon. This is the multiverse. The multiverse is a way of redesigning physical reality, the structure of the vacuum, and the history of the universe.  Some people associate the multiverse with a metaphysical concept but, on the contrary, the concept is deeply rooted in physical reality. 

Compare ourselves with an imaginary astronomer who lives on a planet belonging to a solar system surrounded by a dust cloud impenetrable to electromagnetic radiation. All that the astronomer can observe are the planets and the sun, and he gets obsessed with the idea of computing from fundamental principles the planetary distances because this is all he can see in the universe. Logically he ascribes a fundamental meaning to those numbers because that's all there is to measure. 

But one day a theoretical physicist comes along and tells the astronomer that the universe may be very different from the way it looks. She claims that there are complex structures beyond the dust cloud and that the measurements of planetary distances have to be interpreted in the global context of a larger universe. Hers is not pure philosophy, but it is a precise statement about physical reality. The problem is that it is difficult to measure what's beyond the cloud, but that doesn't make the hypothesis less scientific. It only makes the life of the astronomer harder and forces him to think harder on how to advance knowledge. We are in a similar situation where the cloud dust corresponds to the finite event horizon of our observable universe. 

The multiverse describes a physical reality that challenges the presumption that there must be a single unified theory in the deep UV. In a sense, it is the ultimate Copernican revolution since not even the patch of the universe we live in is special. It implies a revision of the cosmological principle because the universe is approximately homogeneous and isotropic only within our horizon, but may be globally highly non-homogeneous. The multiverse is not an abstract idea, but it is a generic consequence of a large class of inflationary theories, where unavoidable quantum fluctuations of the inflaton spark a chain process with eternal creation of regions that expand faster than the surrounding space.

The multiverse is actually a familiar instrument of our everyday physics toolkit. Take the example of the axion. The starting point is the observation that a SM parameter ($\theta_{\rm QCD}$) should be of order one according to EFT logic, but measurements of the neutron EDM tell us that it is smaller than $10^{-10}$. To deal with the paradox, one promotes the SM parameter to a dynamical field, the axion. During the cosmological history, the axion explores its full field range and takes on all possible values in different patches of the universe. Only after the QCD phase transition, the axion potential creates a non-trivial vacuum structure and dynamics select $\theta_{\rm QCD}=0$ as a preferred point. This is an incarnation of a multiverse explanation for a low-energy parameter. Ultimately, the selection criterion for the axion is symmetry, but the aspect relevant for us is the link with the idea that low-energy parameters may be variables that scan during the cosmological history.

There has been growing interest in the idea that the Higgs mass may not be a rigid parameter but a dynamical variable determined by selection criteria related to the evolution of the universe~\cite{cos1,cos2,cos3,cos4,cos5,cos6,cos7,cos8,cos9,cos10,cos11,cos12,cos13,cos14,cos15,cos16,cos17}. Like every other theorist in the world, I am particularly excited about my own work and I will highlight here an idea~\cite{cos15} developed in this context together with Matthew McCullough and Tevong You. Interestingly, a related but independent framework was recently put forward by Justin Khoury and collaborators~\cite{kho,cos16}.

We are proposing that critical points of quantum phase transitions can become dynamical attractors in the multiverse, thus determining low-energy parameters. Criticality  and not symmetry could be what determines some of the SM parameters. The explanation has a statistical nature, just like in the case of atoms of a gas in thermal equilibrium. If I take one single atom and try to infer its energy from fundamental principles, I don't get very far: any energy is as good as any other. But, if I interpret that atom as part of an ensemble, using statistical equilibrium distributions I can calculate which energy is most probable. Similarly, trying to derive some of the SM parameters from fundamental principles may be futile. If I look at a single universe, any value of those parameters is as good as any other. But, if I interpret the universe as part of a multiverse, I can use statistical distributions and infer that critical points are probabilistically preferred. The change in perspective that we are proposing is to treat the multiverse as a quantum statistical system.

The idea that the Higgs mass may be a dynamical variable determined by cosmology is emerging as a valid alternative to the approach based on the symmetry paradigm. However, we cannot claim any clear winner yet. While the symmetry explanation is under siege by the LHC, the cosmological solution is under attack by an unexpected assailant: quantum gravity. The swampland criteria, in particular the distance and de Sitter conjectures~\cite{swa}, are in conflict with the slow-roll conditions of single-field inflationary models. Moreover, it has been argued on the basis of the $S$-matrix formulation of quantum gravity~\cite{dva} that any efficient cosmological scanning of parameters lies beyond semi-classical calculability and that eternal inflation is inconsistent. These considerations, although still at the level of conjectures,  are a threat for the cosmological approach.

Let me summarise the main point of this discussion about the status of Higgs naturalness after 190 fb$^{-1}$ of LHC data. It is possible to consider alternatives to the traditional symmetry road by relaxing some of the basic hypotheses of naturalness. However, the surprising result is that the consequences of dropping some of the hypotheses on which naturalness rests are even more radical than those of naturalness itself. One is forced to contemplate modifications of QFT, or of the vacuum structure of the universe and its history, leading to a profound change in perspective for fundamental physics. 

Another important development that was stimulated by the negative searches at dark-matter experiments and the LHC is what goes under the name of FIP, which is an umbrella covering a large number of phenomena that involve different kinds of Feebly-Interacting Particles. This activity is having a refreshing and reinvigorating effect on particle physics. It is helping theorists to think outside the box, experimentalists to come up with new techniques borrowed from neighbouring fields, and both theorists and experimentalists to get together and collaborate, as testified by the many recent examples of proposals for radically new types of experiments. In an age in which the time-scale for collider experiments is getting longer and their main target is exploration, particle physics needs a great variety of experimental projects that pursue diversified approaches. 

Theoretical activity in the FIP domain is exploding. However, it would be desirable for this activity not to be driven by attempts to construct models whose only merit is to be experimentally testable. Theoretical particle physics should not forget its basic mission of addressing fundamental questions and advancing in the path of knowledge towards discovering the laws of nature.

The LHC results have changed our perspective on the particle world. The situation is much more complex than it used to be, and the number of logical directions has grown enormously. We have fewer convictions and more untraveled paths. Today, more than ever, we need theoretical research that addresses fundamental conceptual problems with ambitious, wide-ranging goals. We need a bold experimental programme capable of exploring nature at the smallest possible scale. We need a diversified experimental approach that includes large and small scale projects with different goals and  techniques that open bridges with other fields. High-energy colliders remain the most powerful microscope at our disposal to explore nature at small distances and discover the fundamental laws that govern the Universe. Colliders are not single-purpose projects but are, in themselves, a diversified physics programme. This is particularly true today with the extraordinary progress across the precision frontier, which is enabling colliders to cover an ever increasing range of directions in the exploration of fundamental physics.

The world of particle physics is changing, but the world is changing as well. Our responsibility is to provide fundamental knowledge, but to do it responsibly. Particle physics should not simply adapt to societal changes, but lead them. Technologies developed by particle physics are valuable resources for society and must be developed with applications in mind. The global emergency caused by the covid-19 pandemic has taught society the value of science. Particle physics cannot produce vaccines, but has produced the web, which was instrumental for society to survive during the crisis by allowing for the continuation of economical activities, global coordination, communication and social relations. There is much more that particle physics can offer society in the future.

Environmental sustainability is the biggest challenge that our society has to face. We cannot remain indifferent and our way of operating has to change accordingly. As our projects require hundreds of MW, energy consumption, sustainability and efficiency are critical issues for the accelerator complex and computing facilities. Containment of greenhouse gas emissions from detectors must be addressed in every project. As we are an international and dynamical community, we have to reconsider responsibly our travel practices. 

These are interesting times for particle physics: times of great uncertainty, in which our physics perspective is changing, and in which we are laying the foundations for the future of our field. As a community, we must rise to the challenge.


\begin{thebibliography}{99}
\bibitem{Cohen:1998zx}
A.~Cohen, D.~Kaplan and A.~Nelson, Phys.~Rev.~Lett.~82 (1999) 4971, [hep-th/9803132]; A.~Cohen and D.~Kaplan, arXiv:2103.04509.

\bibitem{cos1}
G.~Dvali and A.~Vilenkin, Phys.~Rev.~D 70 (2004) 063501 [hep-th/0304043].

\bibitem{cos2}
G.~Dvali, Phys.~Rev.~D 74 (2006) 025018 [hep-th/0410286].

\bibitem{cos3}
P.~W.~Graham, D.~E.~Kaplan and S.~Rajendran,  Phys.~Rev.~Lett.~115 (2015) 221801 [arXiv:1504.07551].

\bibitem{cos4}
N.~Arkani-Hamed, T.~Cohen, R.~T.~D'Agnolo, A.~Hook, H.~D.~Kim and D.~Pinner, Phys.~Rev.~Lett.~117 (2016) 251801 [arXiv:1607.06821].

\bibitem{cos5}
A.~Arvanitaki, S.~Dimopoulos, V.~Gorbenko, J.~Huang and K.~Van Tilburg, JHEP 05 (2017) 071 [arXiv:1609.06320].

\bibitem{cos6}
A.~Herraez and L.~E.~Ibanez, JHEP 02 (2017) 109 [arXiv:1610.08836].

\bibitem{cos7}
M.~Geller, Y.~Hochberg and E.~Kuflik, Phys.~Rev.~Lett.~122 (2019) 191802 [arXiv:1809.07338].

\bibitem{cos8}
C.~Cheung and P.~Saraswat, arXiv:1811.12390.

\bibitem{cos9}
G.~F.~Giudice, A.~Kehagias and A.~Riotto, JHEP 10 (2019) 199 [arXiv:1907.05370].

\bibitem{cos10}
N.~Kaloper and A.~Westphal, Phys.~Lett.~B 808 (2020) 135616 [arXiv:1907.05837].

\bibitem{cos11}
A.~Strumia and D.~Teresi, Phys.~Rev.~D 101 (2020) 115002 [arXiv:2002.02463].

\bibitem{cos12}
C.~Cs\'aki, R.~T.~D'Agnolo, M.~Geller and A.~Ismail, Phys.~Rev.~Lett.~126 (2021) 091801 [arXiv:2007.14396].

\bibitem{cos13}
N.~Arkani-Hamed, R.~T.~D'Agnolo and H.~D.~Kim, arXiv:2012.04652.

\bibitem{cos14}
G.~F.~Giudice, M.~McCullough and T.~You, arXiv:2105.08617.

\bibitem{cos15}
R.~T.~D'Agnolo and D.~Teresi, arXiv:2106.04591.

\bibitem{cos16}
J.~Khoury and T.~Steingasser, arXiv:2108.09315.

\bibitem{cos17}
V.~Domcke and K.~Schmitz and T.~You, arXiv:2108.11295.

\bibitem{kho}
J.~Khoury and O.~Parrikar, JCAP 1912 (2019) 014 [arXiv:1907.07693];
J.~Khoury, JCAP 06 (2021) 009 [arXiv:1912.06706]; G.~Kartvelishvili, J.~Khoury and A.~Sharma, JCAP 02 (2021) 028 [arXiv:2003.12594]; J.~Khoury and S.~Wong, arXiv:2106.12590.

\bibitem{swa}
H.~Ooguri and C.~Vafa, Adv.~Theor.~Math.~Phys.~21 (2017) 1787 [hep-th/1610.01533]; G.~Obied, H.~Ooguri, L.~Spodyneiko and C.~Vafa, arXiv:1806.08362; P. Agrawal, G. Obied, P. J. Steinhardt and C. Vafa, Phys.~Lett.~B 784 (2018) 271 [arXiv:1806.09718].

\bibitem{dva}
G.~Dvali and C.~Gomez, JCAP 01 (2014) 023 [arXiv:1312.4795]; 
G.~Dvali, C.~Gomez and S.~Zell, JCAP 06 (2017) 028 [arXiv:1701.08776];
G.~Dvali, Symmetry 13 (2020) 3 [arXiv:2012.02133]; G.~Dvali, arXiv:2105.08411.

\end{thebibliography}
\end{document}